\documentclass[aps,prd,showpacs,preprint,groupedaddress]{revtex4-1}

\usepackage[dvips]{graphicx}
\usepackage{dcolumn}
\usepackage{bm}

  \newcommand{\nn}{\nonumber}
  
  \newcommand{\Sura}{\hspace{-1.9mm}\!/}

\begin{document}

\title{Gauge dependence in the anomalous dimension of the gauge invariant canonical decomposition for proton momentum}

\author{Yoshio Kitadono}
 \email{kitadono@impcas.ac.cn}

\author{Pengming Zhang}
 \email{zhpm@impcas.ac.cn}
 \affiliation{Institute of Modern Physics, Chinese Academy of Sciences,\\
Lanzhou, People's Republic of China, 730000.}

\date{\today}

\begin{abstract}
 The gauge dependence in the anomalous dimension of the gauge-invariant-canonical-energy-momentum tensor for proton is studied by the background field method. The naive calculation shows the problem, the absence of the counter term in the gluonic sectors. The analysis shows that the result [Chen {\sl et al.}, Phys.~Rev.~Lett.~{\bf 103},~062001~(2009)] is derived from the background field method after we introduced a trick to avoid the problem except for the gluon-to-gluon sector; it is gauge dependent. The possible reason of this gauge-dependent result comes from the nontrivial treatment of the condition $F^{\mu\nu}_{\rm pure}=0$ at a higher order. This result shows that one needs a further improvement in treating this condition with a covariant way at a higher order by the background field method. In particular, we have to focus on two checkpoints, the gauge independence and zero eigenvalue in the anomalous-dimension matrix, in order to test the validity of the gauge-invariant-canonical-energy-momentum tensor.
\end{abstract}


\maketitle


\section{Introduction}
Attempts to understand the proton spin in terms of quarks and gluons in quantum chromodynamics (QCD) have discussed for long time. The European Muon Collaboration (EMC) in 1987 showed that one cannot explain the proton spin only by quarks in the proton \cite{EMC1,EMC2}. The experiment implied that one should consider the effects not only from the constituent-quark spin, but also from sea quarks, gluons, and corrections at a higher order (see reviews for the spin problem and related topics in Refs. \cite{Review_AEL, Review_HaiYang, Review_Lampe_Reya, Review_Filippone_Ji, Review_Bass, Review_Kuhn_Chen_Leader, Review_Burkardt_Miller_Nowak, Review_Aidala_Bass_Hasch_Mallot} and reference therein). After the experiment, this topic is often called ``spin crisis" or ``spin puzzle" and this is one of challenging problems in QCD even now. 

In the point of view of the quantum field theory, one expects that one will be able to define the operator definition of the contribution from the quark (gluon) spin and quark (gluon)-orbital-angular-momentum (OAM) operator in a gauge-invariant way. Although the expectation value for an operator with a state is more important than the operator itself, a couple of decompositions for the total-angular-momentum operator of quarks and gluons in QCD are proposed. In addition to the spin decomposition, the energy-momentum tensor is closely related to the OAM tensor in the quantum field theory and hence the decomposition of the energy-momentum tensor is sometimes discussed in the context of the spin puzzle. The canonical-energy-momentum tensor in QCD through Noether's theorem does not give a gauge-invariant definition, on the other hand, Belinfante's definition \cite{Belinfante_Rosenfeld} is known as the gauge-invariant expression by adding a surface term to the canonical definition; of course, the surface term does not change the conservation law. 

The canonical-energy-momentum tensor derived from Noether's theorem leads to so-called Jaffe-Manohar decomposition \cite{JM_decomp} and it perfectly separetes the total contribution into the sum of four terms; quark-spin, quark-OAM, gluon-spin, and gluon-OAM with a gauge-variant way. On the other hand, Belinfante decomposition separetes the quark part from the gluon part in a gauge-invariant way, however, each term is not separeted into the spin and OAM term. The author in Ref. \cite{Ji_decomp} considered the further decomposition of the quark sector and it separates the total-quark-angular-momentum term into the sum of the quark-spin and quark-OAM term, and the gluon term is not separated anymore.

In 2008, a new type of gauge-invariant expression for the angumar-momentum and momentum operator was proposed by Chen {\sl et al.} in Ref.~\cite{Chen1} by using split fields for the gluon filed. This result was applied to the anomalous dimension appeared in the rernormalization group equation (RGE) of the energy-momentum tensor for quarks and gluons \cite{Chen2}. This application gives that the momentum fraction carried by gluons in a nucleon at high energies is about one-fifth and this value contradicts to the well-known value, about half, derived by the standard QCD in Ref. \cite{gamma_QCD}. The papers by Chen {\sl et al.} caused intensive debates and a lot of questions. For example, the uniquness of the decomposition, the dependence of Lorentz frame, nonlocarity, and gauge invariance and so on (see recent reviews \cite{Review_Leader, Review_Wakamatsu} on these debates). Currently two different decompositions are known as the gauge invariant expressions, namely, the gauge-invariant-canonical (gic) decomposition and gauge-invariant-kinetic (gik) decomposition. After the paper by Chen {\sl et al.} appeared, the author in Ref.~\cite{Wakamatsu_cova} extended the original decomposition by Chen {\sl et al.} to the covariant form in the four dimension, since the original one was the three dimensional one and hence it is not covariant under Lorentz transformation. The gauge transformations to each split field are same with those to the background field and the quantum field in the background field method (BFM).

The BFM is an alternative way to quantize a field theory and it gives the consistent results with those derived by the standard quantization. Originally the BFM was introduced by DeWitt for the gravity theory \cite{BFM_DeWitt} and applied to gauge theories \cite{BFM_tHooft,  BFM_Grisaru_Nieuwenhuizen_Wu, BFM_Boulware, BFM_Abbott_2loop}. In the BFM , one adds the {\sl background-gauge-fixing term} to the classical Yang-Mills Lagrangian to quantize the theory; this gauge-fixing term is different from the standard one. The BFM splits the gauge field into two pieces, namely, the background (alternatively classical or external) field and the quantum field. The original gauge transformation to the total gauge-field is separated into two transformations. This separation of the gauge transformation is not unique and hence a useful one is chosen. Typically we use the gauge transformations so that the background field transforms like the standard gauge-transformation and the quantum field transforms like the simple rotation in the space of the gauge group. The quantum field is integrated out from the theory in the sense of the path integral and the effective theory after this integration is described by the background field. 

One of the advantages of the BFM is the the manifest gauge-invariance of the theory by using the gauge transformation for the background field. The BFM greatly simplifies loop calculations and typically it is used to calculate the beta function of a given theory. For example, see Ref.~\cite{BFM_Abbott_Review} and the textbook \cite{BFM_Peskin_Schroeder} for calculating the beta function of the non-Abelian gauge theory by using two-different ways, respectively; and see the paper \cite{BFM_Binosi_Quadri} for the relations among the BFM and (anti)BRST symmetry \cite{BRST}, the effective action, and Ward identity. 
The similarities between the gauge transformations for split gauge fields were discussed in Refs.~\cite{Lorce_Path,Lorce_Noether} in terms of a path dependence of Wilison line and Noether's current under the presense of the background gauge field and in Ref.~\cite{Zhang_Pak} in the context of gluon helicity and a little group in the Lorentz group. Acutally the author in Ref.~\cite{Wakamatsu_mom_gamma} evaluated the anomalous dimension of the gic-energy-momentum tensor again and concluded that the results was same with the result in Ref.~\cite{gamma_QCD}. However the author calculated it by own method \cite{Wakamatsu_Feynmanrule}. On the other hand, the authors in Ref.~\cite{Chen2} adopted the Coulomb gauge. Hence, an analysis based on the BFM with a covariant way at the one-loop order was not carried out. 

In Ref.~\cite{Ours.letter}, we studied the one-loop corrections to the anomalous dimensions based on the BFM, because the gic decomposition includes split gauge-fields obeying the same gauge-transformation laws for the background and quantum field in the BFM. The results show that the application of the BFM correctly reproduces two results in the quark sectors of Chen {\sl et al.}'s anomalous-dimension matrix; however, the results in gluonic sectors show the inconsistency in the renormalization for the gluon field in the BFM. Then we considered a trick to overcome this inconsistency and the method led two results: 1) the third result of Chen {\sl et al.}'s  anomalous dimensions is recovered, 2) on the other hand, the fourth result does not coincide with Chen {\sl et al.}'s result. Even worse, the gauge dependence dose not cancels in the result. These results seem to show that the application of the BFM to this problem (at least quark sectors) works well, however, the treatment of the gluonic sector is not perfect. Motivated by our observations, in this paper, we focus on the evaluation of the anomalous dimensions by the BFM, in particular, we compare the anomalous dimensions derived from the Belinfante-improved-energy-momentum tensor by the BFM with those derived from the gic-energy-momentum tensor by the BFM. The analysis in this paper shows that the possible origin of the inconsistency and the imperfect result are from the condition $F^{\mu\nu}_{\rm pure}=0$. In addition, we point out two checkpoints to test the gic decomposition of the energy-momentum tensor; namely, the gauge independence in the gluon-to-gluon sector and the zero eigenvalue in the anomalous-dimension matrix. 

The RGE for the energy-momentum tensor in QCD is briefly reviewed in Sec.~\ref{sec.short.review.RGE}. In particular, the gauge cancellation of the Feynman diagrams and the asymptotic behavior of the momentum fraction carried by gluons in a nucleon at high energies are focused. The same results of the anomalous dimension can be obtained by the BFM in QCD. The gauge cancellation in the Feynman diagrams by the BFM is discussed in Sec.~\ref{sec.short.review.gamma.BFM}. Our analysis of the anomalous dimension based on the gic-energy-momentum tensor by the BFM is showed in Sec.~\ref{sec.gamma.gic} to discuss how Chen {\sl et al.}'s results in the quark sector of the anomalous dimension are derived by the BFM and how gluonic sectors lead to the problems. We consider a trick to overcome these problems in the gluonic sectors. In particular, the detail of comparison of our results with the literature is discussed and two checkpoints for testing the gic decomposition of the energy-momentum tensor are pointed out in Sec.~\ref{sec.discussion}. Last, Sec.~\ref{sec.conclusion} is devoted to the conclusion of this paper and to show a future perspective.

\section{Short review of the RGE of the energy-momentum tensor in QCD \label{sec.short.review.RGE}}   
The asymptotic limit of the momentum fraction carried by gluons in a nucleon at high energies can be predicted by the standard QCD \cite{gamma_QCD}. In this section, we briefly review two topics: 1) how the gauge dependence is canceled and 2) how the asymptotic limit is derived.

\subsubsection{Gauge cancellations in the operator renormalization}
We begin with the definition of Belinfante-improved-energy-momentum tensor:
\begin{eqnarray}
 T^{\mu\nu}_{\rm Bel} &=&  T^{\mu\nu}_{\rm Bel,q} +  T^{\mu\nu}_{\rm Bel,g},\nn\\
 T^{\mu\nu}_{\rm Bel,q} &=& \frac{1}{2}\bar{\psi}\gamma^{\{\mu}iD^{\nu\}}\psi,\hspace{1cm}
  T^{\mu\nu}_{\rm Bel,g} = - \mbox{Tr}\left(F^{\{\mu\alpha}F^{\nu\}}_{\alpha}\right), \label{eq.Bel.Tmunu}
\end{eqnarray}
where the symbol $a^{\{\mu}b^{\nu\}}=a^{\mu}b^{\nu}+a^{\nu}b^{\mu}$ is the symmetrization symbol and the terms including $g^{\mu\nu}$ are ignored since that do not contribute to the three-dimensional-momentum operator of the quark and gluon.
The related Feynman-rules for these operators can be easily derived. For example, the Feynman rules for the quark-quark vertex with a momentum $p$, $V^{\mu\nu}_{\rm qq}(p)$, and for the quark-quark-gluon interaction with the gluon of the Lorentz index $\rho$ and the color $a$, $V^{a;\rho;\mu\nu}_{\rm qqg}$, are given by
\begin{eqnarray}
 V^{\mu\nu}_{\rm qq}(p) &=& \frac{1}{2}\gamma^{\{\mu}p^{\nu\}}, 
 \hspace{1cm}
 V^{a;\rho;\mu\nu}_{\rm qqg} = \frac{g}{2}t^{a}\gamma^{\{\mu}g^{\rho\nu\}}. 
\end{eqnarray}
The Feynman rule for the vertex $A^{a\alpha}(p)A^{b\beta}(-p)$, $V^{ab;\alpha\beta;\mu\nu}_{gg}(p)$, is given by
\begin{eqnarray}
 V^{ab;\alpha\beta;\mu\nu}_{gg}(p)
  &=& \delta^{ab} \left[   p^{\{\mu}p^{\alpha}g^{\beta\nu\}}
                         + p^{\{\mu}p^{\beta}g^{\alpha\nu\}}
                         - p^{\{\mu}p^{\nu\}}g^{\alpha\beta}
                         - p^2 g^{\{\mu\alpha}g^{\beta\nu\}}
                  \right], \label{fig.Feynman.rule.QCD.2g}
\end{eqnarray} 
and the Feynman rule for the $A^{a\alpha}(p_1)A^{b\beta}(p_2)A^{c\gamma}(p_3)$ vertex, $V^{abc;\alpha\beta;\mu\nu}_{gg}(p)$, is given by
\begin{eqnarray}
 V^{abc;\alpha\beta\gamma;\mu\nu}_{ggg}(p_1,p_2,p_3)
  &=& igf^{abc} \left[~  g^{\gamma\{\mu}g^{\beta\alpha} (p_2 - p_1)^{\nu\}}
                       + g^{\beta\{\mu}g^{\alpha\gamma} (p_1 - p_3)^{\nu\}}
                       + g^{\alpha\{\mu}g^{\gamma\beta} (p_3 - p_2)^{\nu\}}
 \right. \nn\\
 &{}& \left. \hspace{1cm}
                       + g^{\beta\{\mu}g^{\alpha\nu\}} (p_2 - p_1)^{\gamma}
                       + g^{\alpha\{\mu}g^{\gamma\nu\}} (p_1 - p_3)^{\beta}
                       + g^{\gamma\{\mu}g^{\beta\nu\}} (p_3 - p_2)^{\alpha}
                  \right],
                  \nn\\ \label{fig.Feynman.rule.QCD.3g}
\end{eqnarray} 
where all momentum is incoming.

\begin{figure}[htb]
  \begin{center}
    \def\SCALEa{0.4}
    \def\SCALEb{0.4}
    \def\SCALEc{0.4}
    \def\SCALEd{0.4}
    \def\OFFSET{10pt}
    \begin{tabular}{cc}
     \hspace{\OFFSET}
      \includegraphics[scale=\SCALEa]{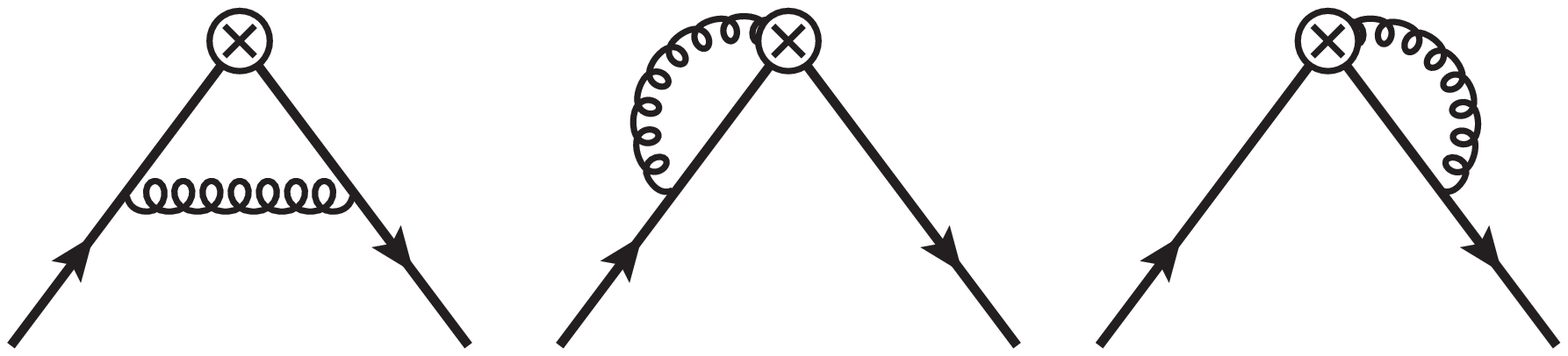} &
     \hspace{\OFFSET}
      \includegraphics[scale=\SCALEb]{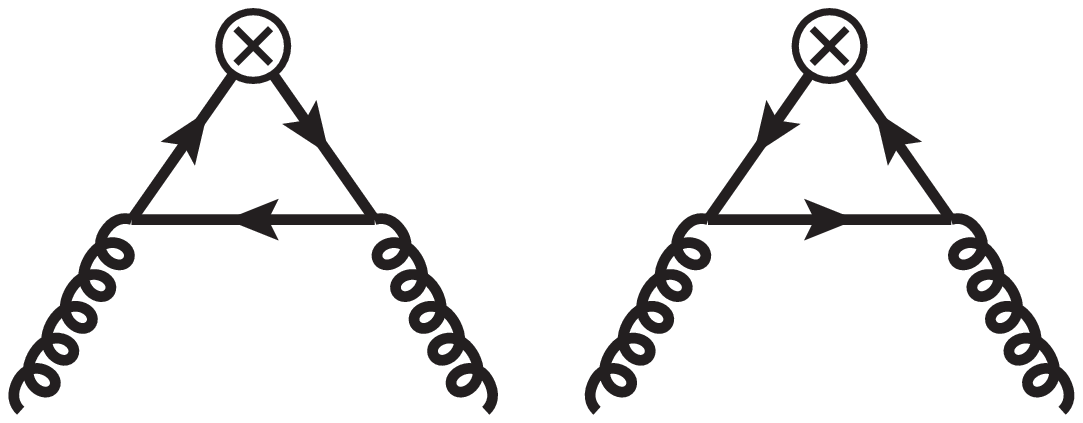} \\
     \hspace{\OFFSET} (a) & \hspace{\OFFSET} (b) \\
      \includegraphics[scale=\SCALEc]{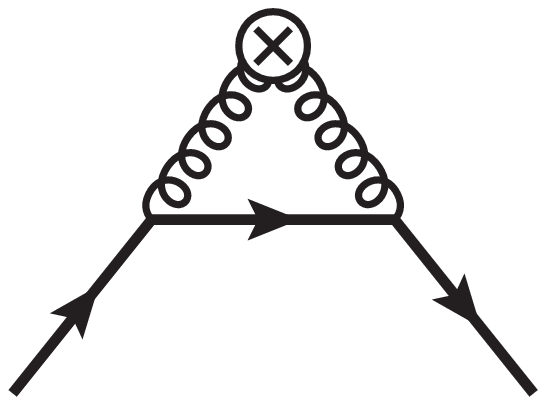} &
     \hspace{\OFFSET}
      \includegraphics[scale=\SCALEd]{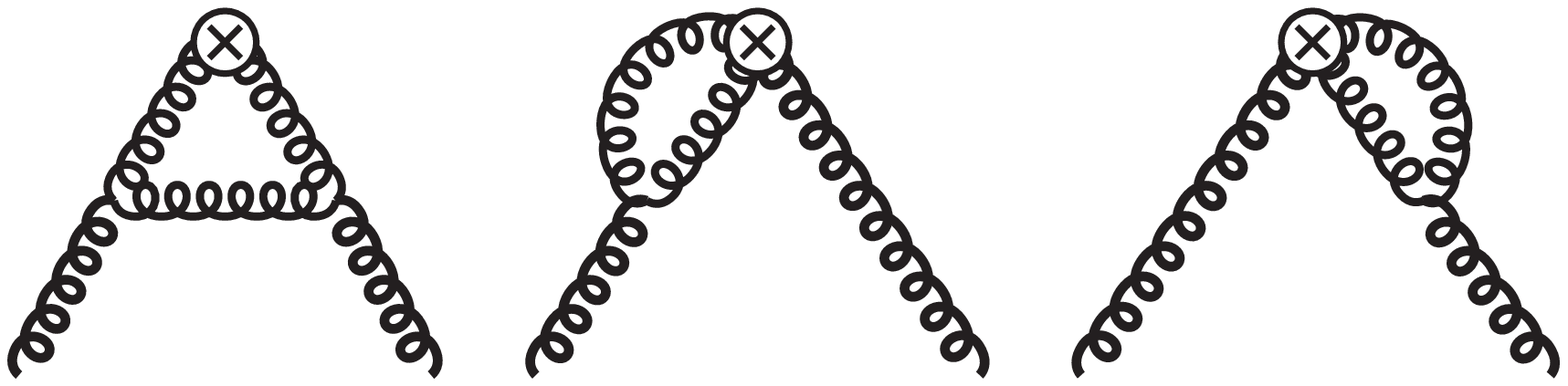} \\
      \hspace{\OFFSET} (c) & \hspace{\OFFSET} (d)
    \end{tabular}
   \caption{One-loop diagrams contributing to the anomalous dimension of the energy-momentum tensor for quarks and gluons in the standard QCD: (a)~$Z^{\rm QCD}_{qq}~(\gamma^{\rm QCD}_{qq})$, (b)~$Z^{\rm QCD}_{gq}~(\gamma^{\rm QCD}_{qg})$, (c)~$Z^{\rm QCD}_{qg}~(\gamma^{\rm QCD}_{gq})$, and (d)~$Z^{\rm QCD}_{gg}~(\gamma^{\rm QCD}_{gg})$. The contributions from the field renormalization should be added to (a) and (d). The symmetric factors to the second and third graph for (d) are $1/2$. }
    \label{fig.Feynman.diagram.QCD}
  \end{center}
\end{figure}

To evaluate the anomalous dimension for the energy-momentum tensor in Eq.~(\ref{eq.Bel.Tmunu}) by the standard QCD, we have to extract divergences of the Feynman diagrams in Fig.~\ref{fig.Feynman.diagram.QCD}(a),~(b),~(c), and (d). Basically we focus on the one-particle-irreducible part of the diagrams without external lines; however the external lines are assumed to combine with the one-particle-irreducible part.
The renormalization constants for the energy-momentum tensor evaluated by the dimensional regularization with the dimension $D=4-2\epsilon$ \cite{Dimreg} and with the modified-minimal-subtraction ($\overline{\rm{MS}}$) scheme \cite{MSbar} for the related diagrams are given in:
\begin{eqnarray}
  Z^{\rm QCD}_{qq} &=& 1 + \frac{g^2}{(4\pi)^2} \left(\frac{8}{3}C_{F}\right)\frac{1}{\bar{\epsilon}},\hspace{1cm}
  Z^{\rm QCD}_{gq} = \frac{g^2}{(4\pi)^2} \left(-\frac{4}{3}T_{R}\right)\frac{1}{\bar{\epsilon}},\nn\\
  Z^{\rm QCD}_{qg} &=& \frac{g^2}{(4\pi)^2} \left(-\frac{8}{3}C_{F}\right)\frac{1}{\bar{\epsilon}}, \hspace{1cm} 
  Z^{\rm QCD}_{gg} = 1 + \frac{g^2}{(4\pi)^2} \left( \frac{4}{3} T_{R}\right)\frac{1}{\bar{\epsilon}}, \label{eq.QCD.Zij}
\end{eqnarray}
with the notation $1/\bar{\epsilon}=1/\epsilon - \gamma_{E} + \ln(4\pi)$, $C_{F}=(N^2_c-1)/(2N_c)$, and $T_{R}=1/2$ for $\mbox{SU}(N_c)$ gauge group. In the point of view of the cancellation of the gauge parameter $\xi$ in the standard-covariant-gauge-fixing procedure, the gauge dependence in $Z^{\rm QCD}_{qg}$ vanishes in the own diagram and $Z^{\rm QCD}_{gq}$ has no gauge dependence due to the absence of gluon propagators, on the other hand, the gauge dependence in $Z^{\rm QCD}_{qq}, Z^{\rm QCD}_{gg}$ cancels between the related diagrams and the field renormalization constants for the quark and gluon respectively:
\begin{eqnarray}
 Z^{\rm QCD}_{2} &=& 1-\frac{g^2}{(4\pi)^2}C_{F}
                     \frac{\xi}{\bar{\epsilon}},\hspace{1cm}
 Z^{\rm QCD}_{3} = 1+\frac{g^2}{(4\pi)^2}\left[\left(\frac{13}{6} - \frac{\xi}{2}\right)C_{G}-\frac{4}{3}n_f T_{R}\right] 
\frac{1}{\bar{\epsilon}},
\end{eqnarray} 
with the definition of the group factors $C_{G}=N_c$, a number of quark flavor, $n_f$. We set $n_f=1$ in computing $Z^{\rm QCD}_{ij}$ in Eq.~(\ref{eq.QCD.Zij}). The diagrams in Fig.~\ref{fig.Feynman.diagram.QCD}(d) are proportional to the group factor $C_{G}$, however, the final result of $\gamma^{\rm QCD}_{gg}$ is proportional to $T_R$. This is because of the cancellation of $C_{G}$ between the related diagrams and the same contribution from $Z^{\rm QCD}_{3}$ and hence only the fermionic contribution of $T_R$ remains in the final result.

By using the operator renormalization, $T^{\rm R,\mu\nu}_{i}=Z^{\rm QCD}_{ij}T^{\rm B,\mu\nu}_{j}$, for the bare (renormalized) operator $T^{\rm B(R),\mu\nu}_{i=q,g}$, we can derive the following RGE:
\begin{eqnarray}
 \frac{d}{d\ln\mu}
 \left(
 \begin{array}{c}
    T^{\rm R,\mu\nu}_{q}(\mu) \\
    T^{\rm R,\mu\nu}_{g}(\mu) 
 \end{array}
 \right)
 &=& -\gamma^{\rm QCD}(\mu)  \left(
 \begin{array}{c}
    T^{\rm R,\mu\nu}_{q}(\mu) \\
    T^{\rm R,\mu\nu}_{g}(\mu) 
 \end{array}
 \right), \label{eq.RGE.Tmunu.QCD.nf1}
\end{eqnarray} 
where the $\gamma^{\rm QCD}(\mu)$ is the anomalous-dimension matrix in the standard QCD. The matrix element $\gamma_{ij}$ is obtained through $Z_{ij}$ by the definition $ \gamma_{ij} \equiv Z_{jk}\left( dZ^{-1}/d\ln\mu \right)_{ki}$ and then the result is given in
\begin{eqnarray}
 \gamma^{\rm QCD}(\mu)
  &\equiv&  \frac{\alpha_s(\mu)}{4\pi}
  \left(
 \begin{array}{cc}
     A &  -B \\
    -A &   B
 \end{array}
 \right),
\end{eqnarray}
with $A=16C_{F}/3, B=8T_{R}/3$. Note that our $\gamma_{ij} $ is transposed to the results in Ref.~\cite{gamma_QCD}. 

\subsubsection{Asymptotic limit for momentum fraction carried by gluons}
If there are $n_f$-flavor quarks, the RGE and anomalous dimension are extended to the following $(n_f+1)\times(n_f+1)$ matrix:
\begin{eqnarray}
\frac{d}{d\ln\mu}
\left( 
\begin{array}{c}
       T^{\rm R,\mu\nu}_{q_1}(\mu) \\
       T^{\rm R,\mu\nu}_{q_2}(\mu) \\
       \vdots \\
       T^{\rm R,\mu\nu}_{q_{n_f}}(\mu) \\
       T^{\rm R,\mu\nu}_{g}(\mu) 
 \end{array}
 \right)
 &=& 
 - \frac{\alpha_s(\mu)}{4\pi}
  \left(
 \begin{array}{ccccc}
      A &      0 & \cdots &      0 & - B \\
      0 &      A & \cdots &      0 & - B \\
 \vdots & \vdots & \ddots & \vdots & \vdots \\
      0 &      0 & \cdots &      A & -B \\
     -A &     -A & \cdots &     -A &  n_f B
 \end{array}
 \right)
 \left(
 \begin{array}{c}
       T^{\rm R,\mu\nu}_{q_1}(\mu) \\
       T^{\rm R,\mu\nu}_{q_2}(\mu) \\
       \vdots \\
       T^{\rm R,\mu\nu}_{q_{n_f}}(\mu) \\
       T^{\rm R,\mu\nu}_{g}(\mu) 
 \end{array}
 \right), \label{eq.RGE.Tmunu.QCD}
\end{eqnarray}
where the $n_f$ dependence in the above equation comes from the flavor dependence in the field renormalization $Z^{\rm QCD}_{3}$ for the gluon field and there is no mixing among different flavors at this order in QCD. We can solve Eq.~(\ref{eq.RGE.Tmunu.QCD}) by introducing the singlet (S) and nonsinglet (NS) bases with a certain initial condition. These bases are defined by
\begin{eqnarray} 
 T^{\rm R,\mu\nu}_{\rm S}(\mu) 
   &\equiv& \sum_{i=1}^{n_f} T^{\rm R,\mu\nu}_{q_i}(\mu),
\hspace{1cm}
 T^{\rm R,\mu\nu}_{\rm NS}(\mu) 
   \equiv   T^{\rm R,\mu\nu}_{q_i}(\mu)
              - \frac{1}{n_f}\sum_{i=1}^{n_f} T^{\rm R,\mu\nu}_{q_i}(\mu),
\end{eqnarray}
then the RGE is reduced the to the simpler form:
\begin{eqnarray}
 \frac{d}{d\ln\mu}
\left( 
\begin{array}{c}
       T^{\rm R,\mu\nu}_{\rm S}(\mu) \\
       T^{\rm R,\mu\nu}_{g}(\mu) 
 \end{array}
 \right)
 &=& 
 - \frac{\alpha_s(\mu)}{4\pi}
  \left(
 \begin{array}{ccccc}
      A &  - n_f B \\
    - A &    n_f B
 \end{array}
 \right)
 \left(
 \begin{array}{c}
       T^{\rm R,\mu\nu}_{\rm S}(\mu) \\
       T^{\rm R,\mu\nu}_{g}(\mu) 
 \end{array}
 \right), \nn\\
  \frac{d}{d\ln\mu} T^{\rm R,\mu\nu}_{\rm NS}(\mu) 
 &=&
 - \frac{\alpha_s(\mu)}{4\pi} A
       T^{\rm R,\mu\nu}_{\rm NS}(\mu).
\end{eqnarray}
The solution to the above matrix-equation can be evaluated by the diagonalization and one can obtain the asymptotic limit by taking $\mu\to\infty$ in the solution.
The asymptotic limit to the solutions for the singlet and gluon are reduced to the forms:
\begin{eqnarray}
 \lim_{\mu \to \infty}T^{\rm R,\mu\nu}_{\rm S}(\mu) 
  &=& \frac{n_f B}{A + n_f B} T^{\rm R,\mu\nu}_{\rm tot} + \cdots 
\hspace{1cm}
  \lim_{\mu \to \infty}T^{\rm R,\mu\nu}_{g}(\mu) 
  = \frac{A}{A + n_f B} T^{\rm R,\mu\nu}_{\rm tot} + \cdots, 
\end{eqnarray}
where the terms described by ``$\cdots$" correspond to the power-suppressed terms when $\mu \to \infty$, the nonsinglet solution is also suppressed by a power in this limit, and $T^{\rm R,\mu\nu}_{\rm tot}=T^{\rm R,\mu\nu}_{\rm S}(\mu)+T^{\rm R,\mu\nu}_{g}(\mu)$ is the total-energy-momentum tensor. As we have already mentioned earlier, the standard QCD gives that the momentum fraction carried by gluons in a nucleon at high energies is $A/(A+n_f B)$; it is reduced to $16/(16+3n_f)$ and gives the value about $1/2$ for $n_f=5$. It is worth noting that the total energy-momentum-tensor $T^{\rm R,\mu\nu}_{\rm tot}$ is scale ($\mu$) independent as expected from the conservation law of the energy-momentum and the solutions of the RGE at the one-loop order explicitly show this desired property. 

The anomalous-dimension matrix, equivalently, the divergences in Fig.~\ref{fig.Feynman.diagram.QCD} are responsible for the momentum fraction carried by gluon in the nucleon. Although this matrix describes the mixing between the singlet and gluon sectors, the singlet sector is simply called the quark sector in the literature and hence we call it the quark sector in this paper. Effectively, we can evaluate the gluonic matrix-elements from the Feynman diagrams including a single quark by multiplying $n_f$ factor. To compare our results with the literature, we redefine the anomalous dimension by the following notation:
\begin{eqnarray}
 \gamma(\mu)  &\equiv& 
 - \frac{\alpha_s(\mu)}{4\pi}
  \left(
    \begin{array}{cc}
      \gamma_{qq} & \gamma_{qg}  \\
      \gamma_{gq} & \gamma_{gg}  
    \end{array}
  \right),
\end{eqnarray}
where the overall sign and the loop factor $\alpha_s(\mu)/(4\pi)$ are factored out from the definition of $\gamma_{ij}$. In this notation, the standard QCD results are
\begin{eqnarray}
  \left(
    \begin{array}{cc}
      \gamma^{\rm QCD}_{qq} & \gamma^{\rm QCD}_{qg}  \\
      \gamma^{\rm QCD}_{gq} & \gamma^{\rm QCD}_{gg}  
    \end{array}
  \right)
  =  \left(
    \begin{array}{cc}
      -\frac{8n_g}{9} & \frac{4}{3}n_f  \\
       \frac{8n_g}{9} & -\frac{4}{3}n_f 
    \end{array}
  \right),
\end{eqnarray}
where $n_g=8$ is the number of gluons and $n_f$ is a number of  quark flavors. On the other hand, Chen {\sl et al.}'s results are 
\begin{eqnarray}
  \left(
    \begin{array}{cc}
      \gamma^{\rm Chen}_{qq} & \gamma^{\rm Chen}_{qg}  \\
      \gamma^{\rm Chen}_{gq} & \gamma^{\rm Chen}_{gg}  
    \end{array}
  \right)
  =  \left(
    \begin{array}{cc}
      -\frac{2n_g}{9} & \frac{4}{3}n_f  \\
       \frac{2n_g}{9} & -\frac{4}{3}n_f 
    \end{array}
  \right),
\end{eqnarray}
where the differences between $\gamma^{\rm QCD}_{ij}$ and $\gamma^{\rm Chen}_{ij}$ come from the quark sectors $\{\gamma_{qq},~\gamma_{gq}\}$, that is, $\gamma^{\rm QCD}_{qq(gq)}=4\gamma^{\rm Chen}_{qq(gq)}$.

\section{Anomalous dimension of Belinfante-energy-momentum tensor by the BFM \label{sec.short.review.gamma.BFM}}
We discuss how the standard results of $\gamma^{\rm QCD}_{ij}$ are derived by the BFM in this section. 
The BFM for the anomalous dimension was studied in Ref.~\cite{Morozov}.
We consider the massless-quark QCD by the background field method with the background gauge fixing term and corresponding ghost term:
\begin{eqnarray}
 \mathcal{L} 
 &=& \mathcal{L}_{\rm cl} + \mathcal{L}_{\rm gf+gh}, \nn\\
 \mathcal{L}_{\rm cl} 
 &=& \bar{\psi}iD\Sura\psi - \frac{1}{2}\mbox{Tr}\left( F_{\mu\nu}F^{\mu\nu} \right), \nn\\
 \mathcal{L}_{\rm gf+gh}
 &=& - \frac{1}{\xi} \mbox{Tr}\left[\left( D^{\mu}_{\rm bg} A^{\rm qt}_{\mu}\right)^2\right]
     - 2 \mbox{Tr}\left( \bar{c} D^{\mu}_{\rm bg}D_{\mu} c\right),
\end{eqnarray}
where the gauge field in the field strength $F^{\mu\nu}=\partial^{\mu}A^{\nu}-\partial^{\nu}A^{\mu}-ig[A^{\mu},A^{\nu}]$ is supposed to be decomposed to the form $A^{\mu} = A^{\mu}_{\rm bg} + A^{\mu}_{\rm qt}$, $D^{\mu}_{\rm (bg)}=\partial^{\mu}-ig[A^{\mu}_{\rm (bg)},~~]$ is the covariant derivative with the total(background)-gauge field, $\xi$ is the gauge parameter, and $(\bar{c})c$ is the (anti)ghost field. Although we wrote the gauge-fixing and ghost terms, the ghost term is not directly relevant to the operator mixing discussed now and is only related to the field renormalization of the gluon field by the BFM. 

This BFM leads to the relevant interactions, $\bar{\psi}A_{\rm qt}\psi$ and $A_{\rm qt}A_{\rm qt}A_{\rm bg}$. First, the Feynman rule of the $A^{a\alpha}_{\rm bg}(p_1)A^{b\beta}_{\rm qt}(p_2)A^{c\gamma}_{\rm qt}(p_3)$ vertex from the Lagrangian is given by \cite{BFM_Abbott_Review}:
\begin{eqnarray}
  V^{abc;\alpha\beta\gamma}_{\rm BFM, ggg}(p_1,p_2,p_3)
  &=& -gf^{abc} \left[~ g^{\gamma\beta} (p_3 - p_2)^{\alpha} 
                      + g^{\gamma\alpha} \left(p_1-p_3-\frac{1}{\xi}p_2 \right)^{\beta}   
                      + g^{\alpha\beta} \left(p_2-p_1+\frac{1}{\xi}p_3 \right)^{\gamma}                      
                  \right],\nn\\ \label{eq.Feynman.rule.BFM.3g.vertex}
\end{eqnarray}
where all momentum is incoming and the Feynman rule of $\bar{\psi}A_{qt}\psi$ is the same with the standard one. 

\begin{figure}[htb]
  \begin{center}
    \def\SCALEa{0.4}
    \def\SCALEb{0.4}
    \def\SCALEc{0.4}
    \def\SCALEd{0.4}
    \def\OFFSET{10pt}
    \begin{tabular}{cc}
     \hspace{\OFFSET}
      \includegraphics[scale=\SCALEa]{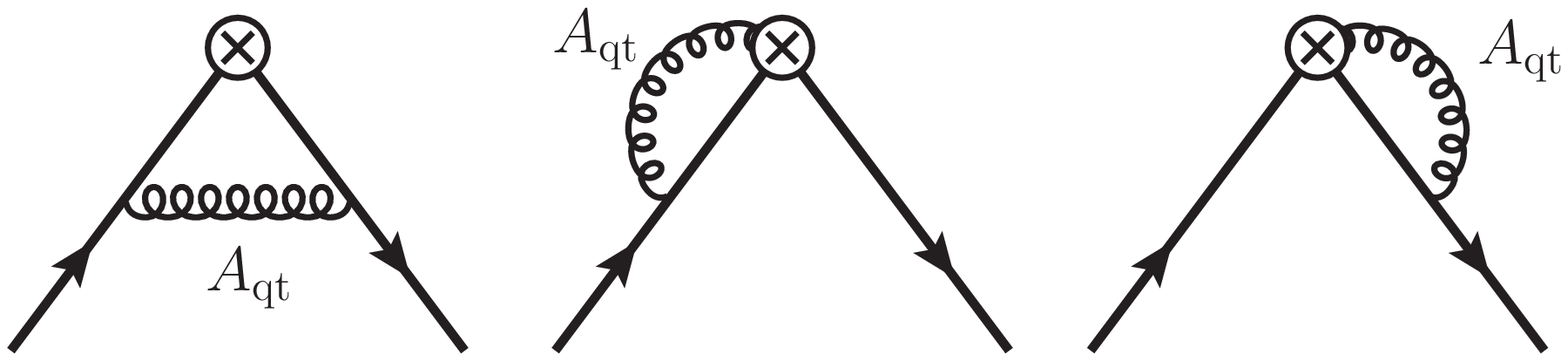} &
     \hspace{\OFFSET}
      \includegraphics[scale=\SCALEb]{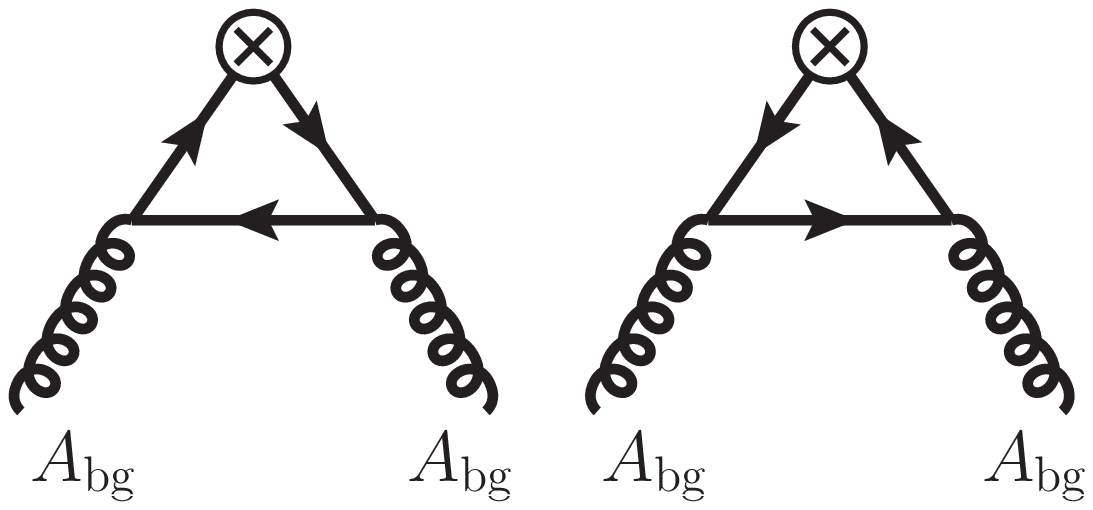} \\
     \hspace{\OFFSET} (a) & \hspace{\OFFSET} (b) \\
      \includegraphics[scale=\SCALEc]{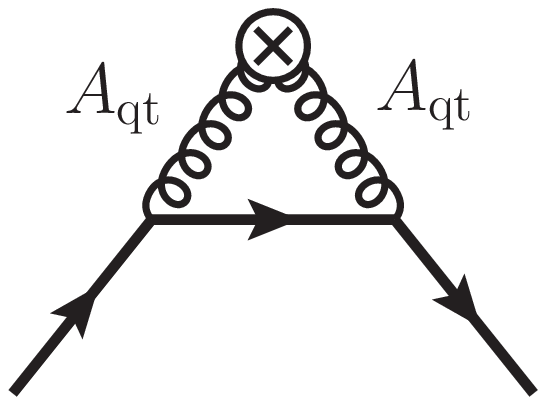} &
     \hspace{\OFFSET}
      \includegraphics[scale=\SCALEd]{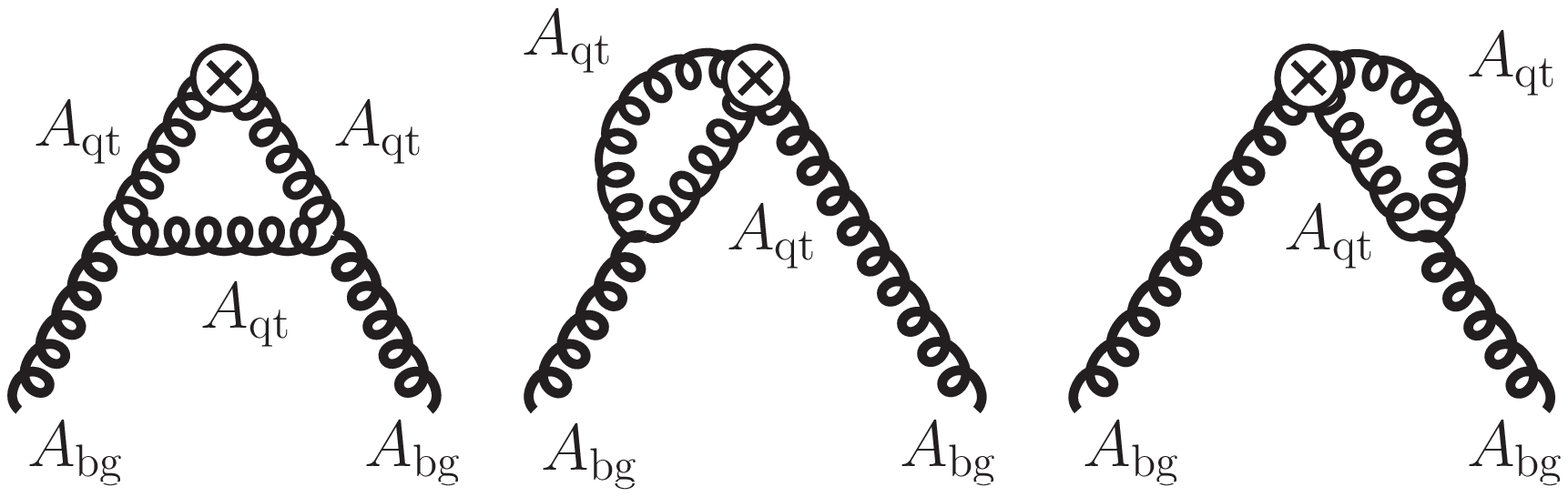} \\
      \hspace{\OFFSET} (c) & \hspace{\OFFSET} (d)
    \end{tabular}
   \caption{One-loop diagrams contributing to the anomalous dimension of the energy-momentum tensor for quarks and gluons in the BFM: (a)~$Z^{\rm BFM}_{qq}~(\gamma^{\rm BFM}_{qq})$, (b)~$Z^{\rm BFM}_{gq}~(\gamma^{\rm BFM}_{qg})$, (c)~$Z^{\rm BFM}_{qg}~(\gamma^{\rm BFM}_{gq})$, and (d)~$Z^{\rm BFM}_{gg}~(\gamma^{\rm BFM}_{gg})$. The contributions from the field renormalization should be added to (a) and (d).}
    \label{fig.Feynman.diagram.BFM}
  \end{center}
\end{figure}

We use the same Belinfante-improved-energy-momentum tensor in Eq.~(\ref{eq.Bel.Tmunu}) by the BFM. Next, we have to consider the modifications of the related rules in the BFM, because two gauge-fields appeared in the BFM change Feynman rules. The Feynman rule for the $A^{a\alpha}_{\rm qt}(p)A^{b\beta}_{\rm qt}(-p)$ vertex from the energy-momentum tensor is same with the Feynman rule for the standard one, namely,
\begin{eqnarray}
 V^{ab;\alpha\beta;\mu\nu}_{\rm BFM, gg}(p) 
 &=& V^{ab;\alpha\beta;\mu\nu}_{gg}(p).  \label{eq.Feynman.rule.BFM.2g}
\end{eqnarray}
The Feynman rule for the $A^{a\alpha}_{\rm qt}(p_1)A^{b\beta}_{\rm qt}(p_2)A^{c\gamma}_{\rm bg}(p_3)$ vertex from the energy-momentum tensor with momenta $(p_1,p_2,p_3)$, colors $(a,b,c)$, and Lorentz indexes $(\alpha,\beta,\gamma)$, $V^{abc;\alpha\beta;\mu\nu}_{\rm BFM, ggg}(p_1,p_2,p_3)$, is given by
\begin{eqnarray}
 V^{abc;\alpha\beta\gamma;\mu\nu}_{\rm BFM, ggg}(p_1,p_2,p_3)
  &=& igf^{abc} \left[~ p^{\{\mu}_{3} g^{\gamma\beta}g^{\alpha\nu\}} 
                      - p^{\{\mu}_{3} g^{\gamma\alpha}g^{\beta\nu\}}
                      + p^{\alpha}_{3} g^{\gamma\{\mu}g^{\beta\nu\}}
                      - p^{\beta}_{3} g^{\gamma\{\mu}g^{\alpha\nu\}} 
 \right. \nn\\
 &{}& \left. \hspace{1cm}
                      + p^{\{\mu}_{2} g^{\gamma\nu\}}g^{\alpha\beta} 
                      - p^{\{\mu}_{1} g^{\gamma\nu\}}g^{\alpha\beta}
                      + p^{\gamma}_{2} g^{\beta\{\mu}g^{\alpha\nu\}} 
                      - p^{\gamma}_{1} g^{\alpha\{\mu}g^{\beta\nu\}}
                  \right], \label{eq.Feynman.rule.BFM.3g}
\end{eqnarray} 
where the all momentum is incoming.
To evaluate the anomalous dimension for the energy-momentum tensor in Eq.~(\ref{eq.Bel.Tmunu}) by the BFM, we have to extract divergences of the Feynman diagrams in Fig.~\ref{fig.Feynman.diagram.BFM}(a),~(b),~(c), and (d).
We can obtain the same results with the standard QCD for the renormalization factors $Z^{\rm QCD}_{ij}$, as expected. For example, we can easily understand that the BFM derives the same results for Fig.~\ref{fig.Feynman.diagram.BFM}(a), (b), (c), because the BFM does not change related Feynman rules. 
However, the loop structure of Fig.~\ref{fig.Feynman.diagram.BFM}(d) is different from that of the standard one; in particular, the cancellation of the gauge parameter by the BFM is different from that by the standard QCD. The rernomalization factor for the background gluon field, $Z^{\rm BFM}_{3}$, is different from $Z^{\rm QCD}_{3}$, because the BFM leads to the relation $Z_{g}\sqrt{Z_{3}^{\rm BFM}}=1$ between $Z^{\rm BFM}_{3}$ and the rernomalization constant for the gauge coupling, $Z_{g}$ \cite{BFM_Abbott_Review}. Hence the following relation holds at the one-loop order,
\begin{eqnarray}
 Z^{\rm BFM}_{3} 
 &=& 1 + \frac{g^2}{(4\pi)^2}\frac{1}{\bar{\epsilon}} \beta_0,
 \hspace{1cm}
\beta_{0} = \frac{11}{3}C_{G} - \frac{4}{3}n_f T_{R},
\end{eqnarray} 
where this relation correctly reproduces the well-known QCD beta-function, $\beta(g)=-\beta_0 g^3/(4\pi)^2$ \cite{beta0}.

The above relation of $Z^{\rm BFM}_{3}$ implies that the gauge dependence must cancels among three diagrams in Fig.~\ref{fig.Feynman.diagram.BFM}(d), because there is no gauge dependence in $Z^{\rm BFM}_{3}$. Furthermore, the symmetric factors to the second and third diagrams in Fig.~\ref{fig.Feynman.diagram.BFM}(d) can be different from the value, $1/2$, in the standard QCD, because the contraction structure of quantum fields in the BFM is different from that in the standard QCD. Let the symmetric factor be $k$ and then the divergence of Fig.~\ref{fig.Feynman.diagram.BFM}(d) can be evaluated by {\small \sc PACKAGE-X} \cite{PackageX} in the following form:
\begin{eqnarray}
\mbox{div.~of Fig.\ref{fig.Feynman.diagram.BFM}(d)} 
 &=& - \frac{g^2}{(4\pi)^2} \frac{C_G}{3} 
       \left[ \left(4k-3\right)\xi + 20k-4\right] 
       \frac{1}{\bar{\epsilon}}
        V^{ab;\alpha\beta;\mu\nu}_{\rm BFM,gg}(p), \label{eq.symm.factor}
\end{eqnarray}
where $\xi$ is the gauge parameter and $V^{ab;\alpha\beta;\mu\nu}_{\rm BFM,gg}(p)$ is the LO tensor structure coming from the $A_{\rm bg}A_{\rm bg}$ interaction. We can easily show that the factor $k=3/4$ not only cancels the gauge dependence, but also reproduces the $Z^{\rm QCD}_{gg}$ in the standard QCD by taking into account the renormalization constant for the gluon field, $Z^{\rm BFM}_{3}$. This cancellation is nontrivial, because the gauge dependence in the BFM is more complicated than the standard one; the gauge dependencies in the BFM appear in the three-gluon vertex of $A_{\rm qt}A_{\rm qt}A_{\rm bg}$ and in the gluon propagators. 
 
Finally, we obtain the expected results,
\begin{eqnarray}
 Z^{\rm BFM}_{qq} &=& Z^{\rm QCD}_{qq}, \hspace{1cm}
 Z^{\rm BFM}_{gq}  =  Z^{\rm QCD}_{gq}, \nn\\
 Z^{\rm BFM}_{qg} &=& Z^{\rm QCD}_{qg}, \hspace{1cm}
 Z^{\rm BFM}_{gg}  =  Z^{\rm QCD}_{gg},
\end{eqnarray}
and therefore the BFM reproduces the result, $\gamma^{\rm BFM}_{ij}=\gamma^{\rm QCD}_{ij}$.

\section{Anomalous dimensions of the gauge-invariant-canonical-energy-momentum tensor by the BFM \label{sec.gamma.gic}}
We consider the anomalous dimensions of the gic-energy-momentum tensor in this section. First, we naively apply the BFM to the gic-energy-momentum and see how the results of the quark sectors in Chen {\sl et al.}'s results are derived by this method and how we encounter the problem of the renormalization in the gluonic sectors \cite{Ours.letter}. Next, we discuss an adhoc method to avoid this problem.

\subsection{Naive application of the BFM to the gic decomposition}
We begin with the gic-energy-momentum tensor, $T^{\mu\nu}_{\rm gic}$, in Ref.~\cite{Wakamatsu_mom_gamma}:
\begin{eqnarray}
T^{\mu\nu}_{\rm gic} 
&=&  T^{\mu\nu}_{\rm gic,q} +  T^{\mu\nu}_{\rm gic,g},\nn\\
 T^{\mu\nu}_{\rm gic,q} &=& \frac{1}{2}\bar{\psi}\gamma^{\{\mu}iD^{\nu\}}_{\rm pure}\psi,\hspace{1cm}
  T^{\mu\nu}_{\rm gic,g} = - \mbox{Tr}\left(F^{\{\mu\alpha} D^{\nu\}}_{\rm pure}A^{\rm phys}_{\alpha} \right), \label{eq.gic.Tmunu}
\end{eqnarray}
where this gic-definition is related to the Belinfante-improved-energy-momentum tensor $T^{\mu\nu}_{\rm Bel}$ through the surface term,
\begin{eqnarray}
 T^{\mu\nu}_{\rm gic} &=& T^{\mu\nu}_{\rm Bel} - \partial_{\alpha} \mbox{Tr}\left( F^{\{\mu\alpha}A^{\nu\}}_{\rm phys}\right), 
\end{eqnarray}
alternatively, each term of the gic-definition is related to those of the Belinfante-definition as:
\begin{eqnarray}
 T^{\mu\nu}_{\rm gic,q} 
 &=& T^{\mu\nu}_{\rm Bel,q} - \frac{g}{2} \bar{\psi} \gamma^{\{\mu}A^{\nu\}}_{\rm phys} \psi, \nn\\
  T^{\mu\nu}_{\rm gic,g} 
 &=& T^{\mu\nu}_{\rm Bel,g} - \mbox{Tr} \left[ F^{\{\mu\alpha } D_{\alpha}       A^{\nu\}}_{\rm phys} \right],
\end{eqnarray}
where both second terms in the above equations are rewritten from the contribution from the surface term by the equation of the motion for the gluon field.

We can derive the modified Feynman-rules to calculate the anomalous dimension based on the gic-energy-momentum tensor. The Feynman rule for $A^{a\alpha}_{\rm phys}(p) A^{b\beta}_{\rm phys}(-p)$ vertex with a momentum $p$, the colors $(a,b)$, and the Lorentz indexes $(\alpha,\beta)$, $V^{ab;\alpha\beta;\mu\nu}_{\rm gic,gg}(p)$ is given by
\begin{eqnarray}
 V^{ab;\alpha\beta;\mu\nu}_{\rm gic,gg}(p)
 &=& \frac{\delta^{ab}}{2} 
     \left[ - 2 g^{\alpha\beta}p^{\{\mu}p^{\nu\}}
            + g^{\alpha\{\mu} p^{\nu\}} p^{\beta} 
            + g^{\beta\{\mu} p^{\nu\}} p^{\alpha}
     \right] \label{eq.Feynman.rule.gic.2g},
\end{eqnarray}
and the Feynman rule for $A^{a\alpha}_{\rm phys}(p_1)A^{b\beta}_{\rm phys}(p_2)A^{c\gamma}_{\rm pure}(p_3)$ vertex from the energy-momentum tensor with a momentum $p$, the colors $(a,b)$, and the Lorentz indexes $(\alpha,\beta)$, $V^{ab;\alpha\beta;\mu\nu}_{\rm gic,gg}(p)$ is given by
\begin{eqnarray}
 V^{abc;\alpha\beta\gamma;\mu\nu}_{\rm gic,ggg}(p_1,p_2,p_3)
 &=& \frac{igf^{abc}}{2} 
     \left[  2 (p_2-p_1)^{\{\mu} g^{\gamma\nu\}}g^{\alpha\beta}
           + p^{\{\mu}_{1}g^{\alpha\gamma}g^{\beta\nu\}}
           - p^{\{\mu}_{2}g^{\beta\gamma}g^{\alpha\nu\}}
\right.\nn\\
&{}& \left. \hspace{1.5cm}
           + p^{\beta}_{1}g^{\alpha\mu\}}g^{\gamma\nu\}}
           - p^{\alpha}_{2}g^{\beta\mu\}}g^{\gamma\nu\}}
     \right],\label{eq.Feynman.rule.gic.3g}\nn\\
\end{eqnarray}
where all momenta is incoming.

\begin{figure}[htb]
  \begin{center}
    \def\SCALEa{0.45}
    \def\SCALEb{0.45}
    \def\SCALEc{0.45}
    \def\SCALEd{0.45}
    \def\OFFSET{40pt}
    \begin{tabular}{cc}
     \hspace{\OFFSET}
      \includegraphics[scale=\SCALEa]{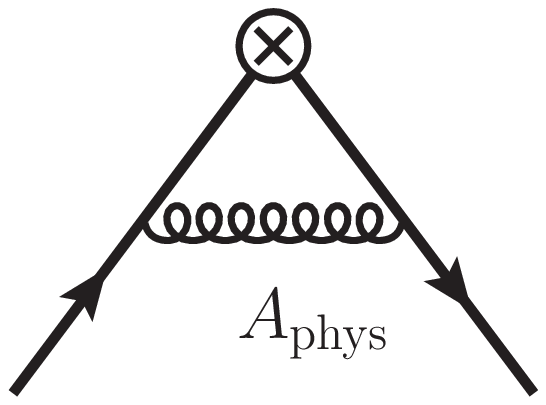} &
     \hspace{\OFFSET}
      \includegraphics[scale=\SCALEb]{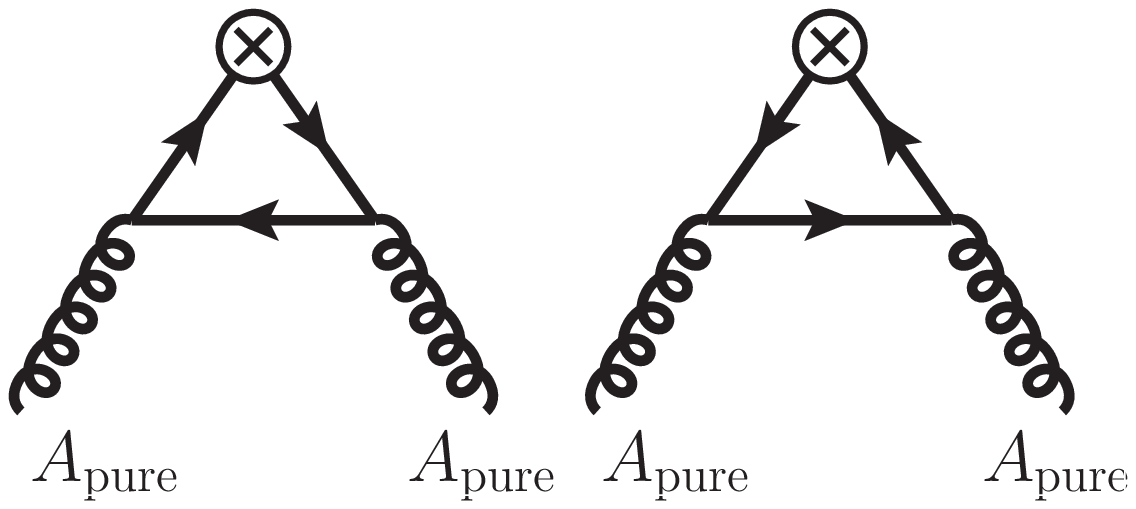} \\
     \hspace{\OFFSET} (a) & \hspace{\OFFSET} (b) \\
      \includegraphics[scale=\SCALEc]{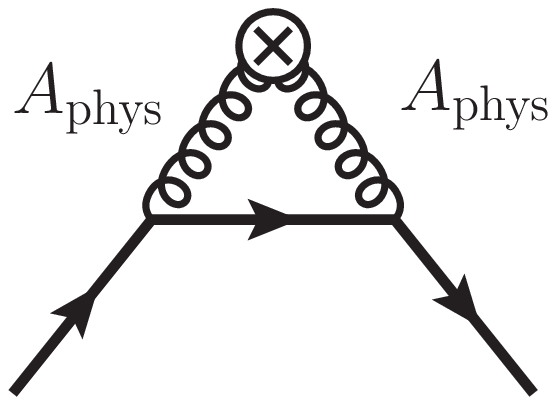} &
     \hspace{\OFFSET}
      \includegraphics[scale=\SCALEd]{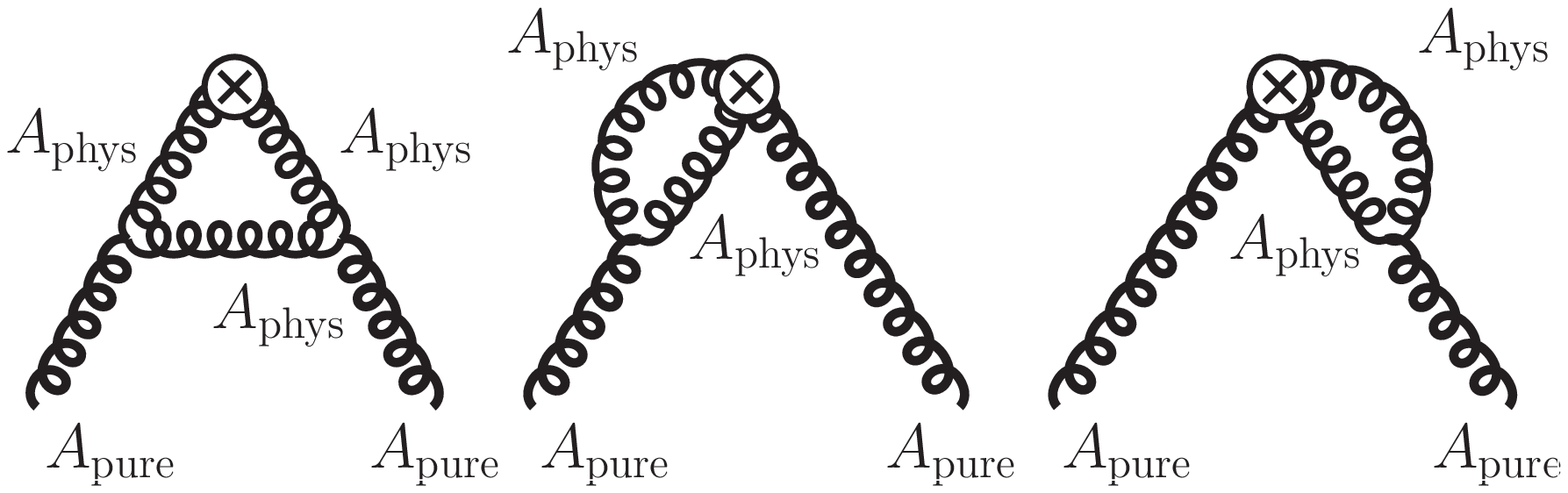} \\
      \hspace{\OFFSET} (c) & \hspace{\OFFSET} (d)
    \end{tabular}
   \caption{One-loop diagrams contributing to the anomalous dimension of the gic-energy-momentum tensor for quarks and gluons in the BFM: (a)~$Z^{\rm gic}_{qq}~(\gamma^{\rm gic}_{qq})$, (b)~$Z^{\rm gic}_{gq}~(\gamma^{\rm gic}_{qg})$, (c)~$Z^{\rm gic}_{qg}~(\gamma^{\rm gic}_{gq})$, and (d)~$Z^{\rm gic}_{gg}~(\gamma^{\rm gic}_{gg})$. The contributions from the field renormalization should be added to (a) and (d).}
    \label{fig.Feynman.diagram.gic.BFM}
  \end{center}
\end{figure}

To evaluate the anomalous dimension for the energy-momentum tensor in Eq.~(\ref{eq.gic.Tmunu}) by the BFM, we have to extract divergences of the Feynman diagrams in Fig.~\ref{fig.Feynman.diagram.gic.BFM}(a),~(b),~(c), and (d).

First, we consider the quark sector $\{\gamma^{\rm gic}_{qq},~\gamma^{\rm gic}_{gq}\}$. It is obvious that Fig.~\ref{fig.Feynman.diagram.gic.BFM}(a) does not have the second and third diagrams of Fig.~\ref{fig.Feynman.diagram.BFM}(a). This difference can be easily understand by the BFM, namely, the background field cannot propagate in loop diagrams in order to keep the gauge invariance of final results. Taking into account this difference, we can reproduce the first Chen {\sl et al.}'s result, $\gamma^{\rm Chen}_{qq}$, namely:
\begin{eqnarray}
 \gamma^{\rm gic}_{qq} &=& \gamma^{\rm Chen}_{qq} = -\frac{2n_g}{9}.
\end{eqnarray}
In addition, the Feynman rule in Eq.~(\ref{eq.Feynman.rule.gic.2g}) is different from the standard rule in Eq.~(\ref{eq.Feynman.rule.BFM.2g}) and hence Fig.~\ref{fig.Feynman.diagram.gic.BFM}(c) reproduces the second Chen {\sl et al.}'s result $\gamma^{\rm Chen}_{gq}$ as
\begin{eqnarray}
  \gamma^{\rm gic}_{gq} &=& \gamma^{\rm Chen}_{gq} = \frac{2n_g}{9}.
\end{eqnarray}

Next, we consider the gluonic sector $\{\gamma^{\rm gic}_{qg},~\gamma^{\rm gic}_{gg}\}$. It is easy to see that Fig.~\ref{fig.Feynman.diagram.gic.BFM}(b) gives the same divergent structure with that in the standard QCD, because the BFM does not change the loop structure like vertexes and fermion propagators in the diagrams. Hence, at first glance, we may conclude the result, $ \gamma^{\rm gic}_{qg} = \gamma^{\rm Chen}_{qg} = 4n_f/3$. However, this is not correct and we encounter the problem at the renormalization of this result in the BFM, namely, the absence of the counter terms for the external lines in Fig.~\ref{fig.Feynman.diagram.gic.BFM}(b). Although we have to renormalize the theory by the external field $A_{\rm pure}$ according to the BFM, we have no such a term in the original definition of $T^{\mu\nu}_{\rm gic}$. Consequently, we cannot remove the divergence and cannot define the anomalous dimension of $\gamma^{\rm gic}_{qg}$ by using the definition in Eq.~(\ref{eq.gic.Tmunu}) in the BFM. Moreover, the same problem appears in the gluon-to-gluon sector, namely, Fig.~\ref{fig.Feynman.diagram.gic.BFM}(d) in computing $\gamma^{\rm gic}_{gg}$.

\subsection{An adhoc method to avoid the problems}
Here we consider an adhoc method to avoid the problem appeared in the previous subsection. One of the reasons why the counter term disappeared is because of the condition $F^{\mu\nu}_{\rm pure}=0$. We have already used this condition in deriving the definition of $T^{\mu\nu}_{\rm gic,g}$. Hence we may try to keep this condition nonzero and set it zero at the end of the calculation. In order to improve the gluonic sector, we go back to the definition of $T^{\mu\nu}_{\rm gic}$ in Eq.~(\ref{eq.gic.Tmunu}) and keep $F^{\mu\nu}_{\rm pure} \neq 0$. Then we find that Eq.(\ref{eq.gic.Tmunu}) should be corrected by the new definition:
\begin{eqnarray}
 T^{\mu\nu}_{\rm gic^{\prime},g} &=& - \mbox{Tr}\left[F^{\{\mu\alpha}(F^{\nu\}}{}_{\rm pure, \alpha} + D^{\nu\}}{}_{\rm pure}A_{\alpha,\rm phys}) \right], \label{eq.gic.prime.Tmunu}
\end{eqnarray}
where the subscript ``gic$^{\prime}$~" stands for the new definition of the gic-energy-momentum tensor.
This new definition does not change the $A_{\rm phys}A_{\rm phys}$ interaction.
By using the new definition, we can recover the gluonic counter term to remove the divergence and it reproduces the third result of Chen {\sl el al.}'s anomalous dimension:
\begin{eqnarray}
 \gamma^{\rm gic^{\prime}}_{qg} = \gamma^{\rm Chen}_{qg} =  \frac{4}{3}n_f T_R.
\end{eqnarray}
To calculate the last piece of the anomalous dimension $\gamma^{\rm gic^{\prime}}_{gg}$, we have to changes Feynman rule of $A_{\rm phys}A_{\rm phys}A_{\rm pure}$ interaction. The modified Feynman rule is given by
\begin{eqnarray}
  V^{abc;\alpha\beta\gamma;\mu\nu}_{\rm gic^{\prime},ggg}(p_1,p_2,p_3)
 &=& igf^{abc}
     \left[ (p_2-p_1)^{\{\mu} g^{\gamma\nu\}}g^{\alpha\beta}
           + \left(p_{1}+\frac{p_2}{2}\right)^{\{\mu}g^{\alpha\gamma}g^{\beta\nu\}}
           - \left(p_{2} + \frac{p_1}{2} \right)^{\{\mu}g^{\beta\gamma}g^{\alpha\nu\}}
\right. \nn\\
&{}& \left. \hspace{1.5cm}
           + \left(p_{1} + \frac{p_2}{2}\right)^{\beta}g^{\alpha\mu\}}g^{\gamma\nu\}}
           - \left(p_{2}+\frac{p_1}{2}\right)^{\alpha}g^{\beta\mu\}}g^{\gamma\nu\}}
     \right].\label{eq.Feynman.rule.gicprime.3g}
\end{eqnarray}
This modified Feynman rule with the same symmetric-factor discussed in Eq.~(\ref{eq.symm.factor}) lead to the following result for $\gamma^{\rm gic^{\prime}}_{gg}$:
\begin{eqnarray}
 \gamma^{\rm gic^{\prime}}_{qg} =  -\frac{4}{3}n_f  -2(3+\xi),
\end{eqnarray}
where this result depends on the gauge parameter. This adhoc method partially works to recover the gluonic counter terms and to reproduce $\gamma^{\rm gic^{\prime}}_{qg}$ and seems to fail in $\gamma^{\rm gic^{\prime}}_{gg}$ due to the additional term.

\section{Discussion \label{sec.discussion}}
First, the naive application of the BFM to the anomalous dimension for the gic-energy-momentum tensor succeeded to reproduce Chen {\sl et al.}'s anomalous dimension correctly, except for the gluonic sectors. Our results contradicted to Ref.~\cite{Wakamatsu_mom_gamma}. The author in Ref.~\cite{Wakamatsu_mom_gamma} developed own method introducing the projection operator to pick up the physical degree of freedom for $A_{\rm phys}$ \cite{Wakamatsu_Feynmanrule}; the autor's conclusion was that the gic-energy-momentum tensor gives the standard results derived by Ref.~\cite{gamma_QCD}. However, in generally speaking, the gic-energy-momentum could give a different result, because the gic definition is different from the Belinfante definition by the surface term. Our explicit calculations based on the BFM give the different Feynman-diagrams and Feynman-rules and finally recovered the most of Chen {\sl et al.}'s anomalous-dimension matrix. We can remember a good example in the Belinfante definition to see how a surface term changes physics. As it is known well that the canonical-energy-momentum tensor derived from Noether's theorem is not always gauge invariant, the Belinfante-improved-energy-momentum tensor derived from the canonical-energy-momentum tensor by adding the surface term is gauge invariant and it gives the gauge independent results at the one-loop order \cite{gamma_QCD} and even at the two-loop order \cite{gamma_QCD_2loop}.

Next, we focus on how our results can be different from the standard results and how Chen {\sl et al.}'s results are derived from the BFM. 
First, we consider $\gamma^{\rm gic}_{qq}$. Comparing Fig.~\ref{fig.Feynman.diagram.BFM}(a) and Fig.~\ref{fig.Feynman.diagram.gic.BFM}(a), it is clear that the gic definition in Eq.~(\ref{eq.gic.Tmunu}) does not generate the second and third diagrams in Fig.~\ref{fig.Feynman.diagram.BFM}(a), because Eq.~(\ref{eq.gic.Tmunu}) does not include the propagating field $A_{\rm phys}$ in the BFM. The absence of the propagating field comes from the surface term, namely, the potential-momentum term. Hence this difference gives our first-result, $\gamma^{\rm gic}_{qq}=\gamma^{\rm Chen}_{qq}$. Second, we consider $\gamma^{\rm gic}_{gq}$ derived from Fig.~\ref{fig.Feynman.diagram.gic.BFM}(c). Although the related Feynman-diagram is same between Fig.~\ref{fig.Feynman.diagram.gic.BFM}(c) and Fig.~\ref{fig.Feynman.diagram.BFM}(c), the related Feynman-rule for $A_{\rm phys}A_{\rm phys}$ interaction is different to each other. This change also comes from the surface term. Hence this difference gives our second-result, $\gamma^{\rm gic}_{gq}=\gamma^{\rm Chen}_{gq}$. Third, on the evaluation $\gamma^{\rm gic^{\prime}}_{qg}$, the loop structures of Fig.~\ref{fig.Feynman.diagram.gic.BFM}(b) is same with that of Fig.~\ref{fig.Feynman.diagram.BFM}(b) and it seems to give the same divergence. Although there was no corresponding counter term to remove this divergence in the original definition in Eq.~(\ref{eq.gic.Tmunu}), now we have the counter term thanks to the new definition in Eq.~(\ref{eq.gic.prime.Tmunu}). The absence of this counter term is related to the treatment of $F^{\mu\nu}_{\rm pure}$ appeared in the surface term. If we keep it  nonzero, we can keep the suitable counter term to reproduce the anomalous dimension $\gamma^{\rm gic^{\prime}}_{qg}$. Hence we can reproduce our third-result, $\gamma^{\rm gic^{\prime}}_{qg}=\gamma^{\rm Chen}_{qg}$. Last, we focus on the anomalous dimension $\gamma^{\rm gic^{\prime}}_{gg}$ in Fig.~\ref{fig.Feynman.diagram.gic.BFM}(d) and in particular how it is different from Fig.~\ref{fig.Feynman.diagram.BFM}(d). Although Fig.~\ref{fig.Feynman.diagram.gic.BFM}(d) is the same with Fig.~\ref{fig.Feynman.diagram.gic.BFM}(d), the Feynman rules in Eqs.~(\ref{eq.Feynman.rule.gic.2g}) and (\ref{eq.Feynman.rule.gicprime.3g}) are different from the standard QCD with the BFM; hence the cancellation of the gauge dependence which works both in the standard method and in the BFM is altered. Actually, we obtain the gauge-dependent result because of this altered gauge-cancellation.

These results mean that the application of the BFM itself will be consistent with the Chen {\sl et al.}'s quark-sector results, however, the adhoc method to improve the gluonic sectors is not enough to reproduce the gluonic results perfectly. Furthermore, the problem of the gluonic term is the counter term and it seems to be related to the condition $F^{\mu\nu}_{\rm pure}=0$. In particular, the gluon-to-gluon sector, $\gamma^{\rm gic^{\prime}}_{gg}$, is sensitive to the way how we treat this condition and our adhoc method lead to the strange result that the ``gauge-invariant-canonical" definition gives the gauge-dependent result. This strange result will be improved, if we can correctly treat the condition of $F^{\mu\nu}_{\rm pure}=0$ at the one-loop order within the BFM.

Most importantly, it is not quite obvious whether we can exactly reproduce Chen {\sl et al.}'s $\gamma^{\rm gic^{\prime}}_{gg}=\gamma^{\rm Chen}_{gg}=-4n_f/3$ or not; namely, we may obtain a gauge-independent result like the form $\gamma^{\rm gic^{\prime}}_{gg}=-4n_f/3+a$ with $a \neq 0$, even after we correctly handle the condition $F^{\mu\nu}_{\rm pure}=0$ at the one-loop order within the BFM by an improved method. This is because that the calculation of the anomalous dimension for the gic-energy-momentum tensor is not explicitly checked by a covariant way within the BFM. Therefore, the gauge independence and the gluonic contribution which is proportional to $C_G$ factor in $\gamma^{\rm gic}_{gg}$ are nontrivial. However, if we obtain the gauge-invariant result like $\gamma^{\rm gic^{\prime}}_{gg}=-4n_f/3 + a$ with $a\neq 0$, then such a solution of the RGE will give a scale-dependent result for the total energy-momentum tensor and cannot be a physical result. This is because that the scale independence in the total-energy-momentum tensor is the consequence of the energy-momentum conservation and it is achieved by the zero eigenvalue in the anomalous-dimension matrix at the one-loop order; the zero eigenvalue requires $a=0$. Hence we expect that the improved method to handle the gluonic sectors of the anomalous dimensions will give the result $\gamma^{\rm gic^{\prime}}_{gg}=-4n_f/3$, if the gic definition is physically correct. In other words, the gic definition of the energy-momentum tensor is unphysical, if $\gamma^{\rm gic^{\prime}}_{gg} \neq -4n_f/3$ with $\gamma^{\rm gic^{\prime}}_{ij} = \gamma^{\rm Chen}_{ij}$ for $(i,j) \neq (g,g)$.  

In any case, one needs more theoretical developments on the treatment of the gluonic terms at a higher order with a covariant way within the BFM. One of the possible way is a method of Lagrange multiplier to handle this condition in the quantum field theory \cite{Bjarke} and we will discuss it in the future publication.

\section{Conclusion \label{sec.conclusion}}
We have studied the anomalous-dimension matrix for the gauge-invariant-canonical definition of the energy-momentum tensor by the background field method. The analysis reproduced Chen {\sl et al.}'s anomalous-dimension matrix except for the gluonic sectors and the reason why we could not derive the gluonic sectors is the absence of the gluonic counter term; namely, the naive gic-definition has the inconsistency in the renormalization in the BFM. This inconsistency comes from the condition $F^{\mu\nu}_{\rm pure}=0$, that is, how we treat this condition in the whole calculation. Then we considered an adhoc method to overcome this problem by keeping $F^{\mu\nu}_{\rm pure}$ nonzero by the end of the all calculation and the anomalous dimension $\gamma^{\rm Chen}_{qg}$ in Chen {\sl et al.}'s result was correctly derived; however, $\gamma^{\rm Chen}_{gg}$ was not derived and the gauge dependence remains in the final result. This means that we could reproduce three-quarter of Chen {\sl et al.}'s results, $\gamma^{\rm Chen}_{qq}$, $\gamma^{\rm Chen}_{gq}$, and $\gamma^{\rm Chen}_{qg}$ by the BFM. 

The current results in the attempt to derive the anomalous dimensions of the gic-energy-momentum tensor by the BFM implies that we have to develop a method to handle the condition $F^{\mu\nu}_{\rm pure}=0$ correctly without losing the renormalizability in the BFM. A possible way will be to introduce a Lagrange multiplier to this condition. In particular, the explicit path integral formalism will be necessary to achieve this purpose. 
Furtheremore, we pointed out {\sl two checkpoints} to test the validity of the gic-energy-momentum tensor; that is, gauge-invariant results for all matrix element of the anomalous dimension and the zero eigenvalue. Although we focused on the renormalization group equation and the anomalous dimension of the gic-energy-momentum tensor, this application of the BFM to the gic decomposition of the total-angular-momentum tensors for the quark and gluon should be considered in the future, because the anomalous dimensions of the total-angular-momentum tensor is related to those of the energy-momentum tensor \cite{Ji_RGE_OAM_qg}. Therefore the complete treatment of the gic and gik decomposition of the energy-momentum tensor based on the BFM is important and it is necessary to check its consistency in the renormalization of the gluonic term at a higher order.

\section*{Acknowledgment}
Y.~K. thanks Sven Bjarke Gudnason, Jarah Evslin, and Emilio Ciuffoli for helpful discussions and thanks Tsuneo Uematsu and Masashi Wakamatsu for useful comments on this work. 
This work is supported in part by National Natural Science Foundation of China under Grant No.~11575254.


\end{document}